# Spreadsheets and the Financial Collapse
3.031

Grenville J. Croll
Chair, European Spreadsheet Risks Interest Group
www.eusprig.org
Grenville@spreadsheetrisks.com

**ABSTRACT**

*We briefly review the well-known risks, weaknesses and limitations of spreadsheets and then introduce some more. We review and slightly extend our previous work on the importance and criticality of spreadsheets in the City of London, introducing the notions of ubiquity, centrality, legality and contagion. We identify the sector of the financial market that we believed in 2005 to be highly dependant on the use of spreadsheets and relate this to its recent catastrophic financial performance. We outline the rôle of spreadsheets in the collapse of the Jamaican banking system in the late 1990's and then review the UK financial regulator's knowledge of the risks of spreadsheets in the contemporary financial system. We summarise the available evidence and suggest that there is a link between the use of spreadsheets and the recent collapse of the global financial system. We provide governments and regulating authorities with some simple recommendations to reduce the risks of continued overdependence on unreliable spreadsheets. We conclude with three fundamental lessons from a century of human error research.*

**1 INTRODUCTION**

Four years ago [Croll, 2005], we interviewed anonymously 23 senior individuals to determine their views regarding the importance and criticality of spreadsheets in the City of London, where they all worked. The details of our motivation, methodology, selection of participants, method of data collection and the background and structure of the interviews are given in that earlier paper. Interviewees provided much useful information, which was best summarized by a financial regulator:

> *"Spreadsheets are integral to the function and operation of the global financial system"*

This succinct comment was reviewed and validated by a senior investment research professional working in a first tier investment bank who added:

> *"Agreed. In addition I would say that the majority of people who use the financial system are not appreciative of the rôle that spreadsheets play"*

Based upon the evidence provided by the interviewees, and the historical evidence of the weaknesses and limitations of spreadsheets, we concluded our article thus:

> *"Spreadsheets have been shown to be fallible, yet they underpin the operation of the financial system. If the uncontrolled use of spreadsheets continues to occur in highly leveraged markets and companies, it is only a matter of time before another 'Black*

*Swan' event occurs [Taleb, 2001], causing catastrophic loss. It is completely within the realms of possibility that a single, large, complex but erroneous spreadsheet could directly cause the accidental loss of a corporation or institution, significantly damaging the City of London's reputation"*

At the time, this conclusion was felt to be rather bold, however following distribution of the paper to the interviewees and during the peer review process, no adverse comments arose and the text was left as written. We are thus in the position of having:

i) An extensive spreadsheets risks research base which identifies spreadsheets as a considerable weakness in organisations
ii) Research evidence outlining the ubiquity, importance and criticality of spreadsheets in the City of London
iii) An *a priori* hypothesis about catastrophic financial loss due the use of untested spreadsheets in the City of London and
iv) A systemically collapsed financial system deeply affecting the City of London

Our present research hypothesis, an extension of iii) above, is to determine by a methodical examination of the evidence available to us whether or not spreadsheets played a rôle in the collapse of the global financial system.

We now review the spreadsheet risks research base and our previous work in more detail, extending the work slightly by adding some email and other evidence that was omitted. We then examine the rôle played by spreadsheets in the collapse of the Jamaican banking system in the late 1990's. Following this, we outline the UK Financial Regulator's knowledge of the problems with spreadsheets by reviewing and summarising a series of public presentations on the issue. We cover briefly the fiduciary duty of UK company directors and their compliance in light of their permitting the extensive use of untested and uncontrolled spreadsheets in financial recording and reporting. Following our conclusions we repeat and then extend a set of simple recommendations.

**2. SPREADSHEET RISKS RESEARCH**

There is a very extensive research base on the risks of using spreadsheets within business [Panko, 2000][Panko & Ordway, 2005][Powell, Baker & Lawson, 2008]. Much of the research has been coordinated and progressed by the European Spreadsheet Risks Interest Group [Chadwick, 2003][EuSpRIG, 2009]. Further significant work improving the end user approach to software has been undertaken by the EUSES consortium [EUSES, 2009]. The main known risks of spreadsheets include:

a) Human Error – To err is human, hence the majority (>90%) of spreadsheets contain errors. Because spreadsheets are rarely tested [Panko, 2006][Pryor, 2004] these errors remain. Recent research has shown that about 50% of spreadsheet models used operationally in large businesses have material defects [Powell, Baker, Lawson, 2007] [Croll, 2008]. Approximately 50% of executives recently surveyed had encountered spreadsheet related problems up to and including staff dismissal [Caulkins, Morrison & Weideman, 2007].

b) Fraud – Because of the ease with which program code and data is mixed, spreadsheets are the perfect environment for perpetrating fraud [Mittermeir, Clermont, Hodnigg, 2005]. The $600m fraud perpetrated by John Rusnak at AIB/Allfirst was spreadsheet related [Butler, 2002]. Other spreadsheet related frauds have occurred and have been notified to the regulator, but have not been reported.
c) Overconfidence – Because spreadsheet users do not go looking for errors, they don't find any or many. Spreadsheet users are therefore overconfident in their use of spreadsheets [Panko, 2003]
d) Interpretation – Translation of a business problem into the spreadsheet domain can *"...lead to a position where decision makers may act in the belief that decisions can be made with confidence on the output from the spreadsheet despite evidence to the contrary"* [Banks & Monday, 2002]
e) Archiving – *"The case of failed Jamaican commercial banks demonstrates how poor archiving can lead to weaknesses in spreadsheet control that contribute to operational risk"* [Lemieux, 2005]. We review this case history of a recent systemic banking collapse in a later section.

Unfolding events suggest that there are several further categories of risks related to spreadsheets that have not been studied at all:

f) Assumptions – Most spreadsheet models rely upon a fairly lengthy series of explicit or more usually, implicit, assumptions. Not least of these is the *Ceteris Paribus* assumption of all other things being (or remaining) equal. Clearly the assumptions underlying many spreadsheet models now no longer apply.
g) Opacity – Initial use of a spreadsheet gives a feeling of transparency, where any and all aspects of the model can be clearly addressed and viewed. With larger models, more complex models, badly structured models or models that have been developed by somebody else, the spreadsheet becomes opaque and unwieldy.
h) Reification – Once information is put into a spreadsheet, it becomes concrete: *"... things changed with the intrusion of the spreadsheet. When you put an Excel spreadsheet into computer-literate hands you get a 'sales projection' effortlessly extending ad infinitum! Once on paper or on a computer screen .....the projection takes on a life of its own, losing its vagueness and abstraction and becoming what philosophers call reified, invested with concreteness; it takes on a new life as a tangible object"* [Taleb, 2007]
i) Enterprise Interoperability – The interconnected heterogeneous nature of the financial system and the spreadsheets contained within it require us to recognise that we can no longer continue with the closed world deterministic assumptions of traditional IT: *"...for with globalisation we are no longer concerned with intra-operability contained within closed systems but with inter-operability between open systems. The closed world assumption of Gödel completeness has been adequate for the logic of closed systems but his undecidable theorems show that the logic of open systems needs to go beyond a mathematics of axioms, sets and number".* [Heather & Rossiter, 2008]. Heather recently expressed the view that this issue underpinned the present crisis [Heather, 2008].

The above points comprise a high level taxonomy of spreadsheet risks. There has been substantive previous work on taxonomies of spreadsheet errors [Rajalingham et al, 2000][Panko, 2008]. We note that previous taxonomies do not address issues relating to the risks involved in the use of large systems of interlinked spreadsheets or some of the other human factors identified above.

## 3. THE RÔLE OF SPREADSHEETS IN THE FINANCIAL SYSTEM

Our previous paper [Croll, 2005] gave a reasonably comprehensive view of how spreadsheets are used in the City of London. More recent direct personal experience in the Enterprise Spreadsheet Management System marketplace [Baxter, 2007] [Saadat, 2007] [Perry, 2008] within the City of London and other global financial centres provided further useful information. We did not find any evidence which contradicted that reported previously. We summarise the various themes developed in our earlier paper, relating them to current events and adding several other aspects.

### 3.1 Ubiquity

Amongst a number of observations highlighting the truly ubiquitous use of spreadsheets in the City of London, we made an observation that:

> *"Spreadsheets are particularly heavily used in the more innovative and hence more recent parts of the market, particularly credit derivatives, an area of the market with a significant daily turnover"*

We note that Collateralised Debt Obligations (CDO's), are a popular form of credit derivative. CDO's offer a structured payoff depending upon default levels of underlying assets (including further CDO's) and have played a significant part in the present financial crisis, earning the sobriquet "Toxic Debt". There is good evidence that CDO's are valued within complex spreadsheets:

> *"Standard & Poor's CDO Evaluator Handbook Version 3.0. An Introduction to and operational overview of Standard & Poor's Evaluator Tool, an automated spreadsheet application used for rapid, complex evaluation of the credit quality of CDO asset portfolios".* [Standard & Poor's, 2006]

Note that we have not examined this spreadsheet, only the published documentation, in the manner of our analysis of some spreadsheet documentation used in clinical medicine [Croll & Butler, 2006].

Recently, the UK operation of a large international investment bank was fined £5.6m for breaches of several regulatory principles by the Financial Services Authority:

> *"2.33.3. The booking structure relied upon by the UK operations of Credit Suisse for the CDO trading business was complex and overly reliant on large spreadsheets with multiple entries. This resulted in a lack of transparency and inhibited the effective supervision, risk management and control of the SCG {Structured Credit Group}"* [FSA, 2008].

The problems with CDO's were first identified in 2007, however the full system collapse did not start until a year later:

> *"....Merrill Lynch & Co.'s threat to sell $800 million of mortgage securities seized from Bear Stearns Cos. hedge funds is sending shudders across Wall Street. A sale would give banks, brokerages and investors the one thing they want to avoid: a real price on the bonds in the fund that could serve as a benchmark. The securities are known as collateralised debt obligations, which exceed $1 trillion and comprise the fastest-growing part of the bond market. Because there is little trading in the securities, prices may not reflect the highest rate of mortgage delinquencies in 13 years. An auction that confirms concerns that CDOs are overvalued may spark a chain reaction of write downs that causes billions of dollars in losses for everyone from hedge funds to pension funds to foreign banks. Bear Stearns, the second-biggest mortgage bond underwriter, also is the biggest broker to hedge funds."* [Pittman, 2007]

Note that since there was very little trading in CDO's, the financial valuations expressed in CDO spreadsheet models were reified in the manner summarised earlier by Taleb.

Note also that spreadsheet use in the Peoples Republic of China may be less ubiquitous than in the West [Turner, 1995] and may provide a useful contrast in future research.

Our previous observations about the ubiquity of spreadsheets and their heavy use in the more recent credit derivatives marketplace are therefore well founded. Complex financial instruments such as CDO's are implemented within, and valued by, large and complex spreadsheets. At least one large firm has been fined by the financial regulator for failure to control these spreadsheets. CDO's and other credit derivatives played a leading rôle in the collapse of the global financial system.

### 3.2 Importance

Recent discussions with the senior management teams at a variety of financial institutions revealed that spreadsheets continue to play a rôle at least as important as that depicted in our earlier paper. We give some examples, though there are very many more:

> a) We were asked to investigate a spreadsheet belonging to a top investment fund that was a member of the Investment Management Association (IMA). The spreadsheet entitled (perhaps) BALDEC08.XLS contained the profit & loss statement & balance sheet of the fund for 31 Dec 08. The Net Asset Value closely matched that on the IMA website. This large & complex spreadsheet had not undergone any formal testing or approval and was undocumented [Payette, 2006], unprotected and uncontrolled.

> b) One very large institution with over 750 business critical spreadsheets in use in the finance and other departments, recently employed in excess of 30 specialist contractors for a substantial period to remediate and validate these spreadsheets prior to placing them under a formal system of control.

> c) A large bank was using a suite of several hundred spreadsheets to underpin the operations of one of its key lending departments. There was no evidence that any formal testing of these spreadsheets had taken place and at the time of writing they were uncontrolled.

d) A trading desk of a large foreign bank used spreadsheets as the basis for its derivatives operations.

The firm in the second example above was one of only a handful of UK organisations known to be systematically and professionally addressing business critical spreadsheets used in financial reporting and other important processes. By contrast, we became aware of some large organisations who were simply placing, or intending to place, untested spreadsheet applications under formal control, without testing or validating them first. One or two very large firms had set up manual spreadsheet control systems, which are known to fall into disrepair within a year or so. One FTSE 100 organisation denied the use of spreadsheets for anything other than innocuous purposes. The denial was less than convincing.

It would appear to be the case that there are presently about 25 departmental installations of Enterprise Spreadsheet Management systems within the City of London. That is, a small fraction of the institutions have begun to acknowledge the importance of spreadsheets and have asserted a need to discover, risk assess, test and control them. Most of these installations are within institutions that became insolvent and were subsequently bailed out by the authorities. It would appear to be the case, at present, that the overwhelming majority of key & critical spreadsheets used within UK based (and by implication, global) financial institutions are completely uncontrolled and that about 50% of them are also untested.

### 3.3 Criticality

We previously defined a critical spreadsheet as one where:

> "*Material error could compromise a government, a regulator, a financial market, or other significant public entity and cause a breach of the law and/or individual or collective fiduciary duty. May place those responsible at significant risk of criminal and/or civil legal proceedings and/or disciplinary action*"

The large number of spreadsheets containing the details and valuations of myriad CDO's and other complex financial instruments have clearly compromised governments, regulators and financial markets. As the magnitude of the present financial crisis has become clear, fingers of blame are beginning to search for causes and those responsible. We believe that spreadsheets, and those who used them, or permitted their use, in ignorance of the well-known risks [e.g. Miller, 2005] and consequences, share much of the blame. Given the risks of using spreadsheets and their ubiquity in financial trading, recording and reporting, it is difficult to see how most organisations comply with the basic tenets of the Companies Act, let alone higher statutes such as Sarbanes Oxley and Basle II [Panko & Ordway, 2005].

### 3.4 Legality

The question of the fiduciary responsibility of company directors and officers in relation to the use of untested and uncontrolled business critical spreadsheets has been initially addressed [Cleary, Norris-Jones, Madahar, 2005]. The general question of fiduciary duty given the present circumstances, is within the current debate [Price, 2009]. The UK

Companies Act is very clear on the requirement that directors shall keep accurate records, as section 221 of the act states that:

*(1) Every company shall keep accounting records which are sufficient to show and explain the company's transactions and are such as to—*

*(2) disclose with reasonable accuracy, at any time, the financial position of the company at that time, and*

*(3) enable the directors to ensure that any balance sheet and profit and loss account prepared under this Part complies with the requirements of this Act*

Clearly, given that approximately half of spreadsheets used in the operational and financial management of large corporates have been found to be materially defective, it would appear to be the case that, in recent years, many directors of UK based banks and financial institutions may not have fully complied with section 221 of the Companies Act. We suggest that it may indeed be unlawful for companies to continue using untested and uncontrolled spreadsheets in critical parts of the financial recording and recording infrastructure. The issue of undue reliance on (spreadsheet based) financial models has now come before the UK courts [Herman, 2008]

Moreover, in our view, fiduciary responsibility in respect of the use and misuse of spreadsheets lies both sides of the regulatory divide.

**3.4 Centrality**

The FSA's Gabriel and Gabriella [FSA, 2009] systems for collecting periodic (i.e. daily, weekly, monthly & annual) statutory financial information from regulated institutions are spreadsheet based. The Bank of England likewise routinely collects financial information from across the markets on spreadsheets. The daily cash reconciliations performed by the clearing banks after the 15:30 daily close are performed on spreadsheets [BCS, 2000]. There exists within the City of London a critical spreadsheet called LIBOR.XLS. The capital adequacy ratios of the institutions were known to be stored and monitored from within a regulator's spreadsheet.

The regulating authorities use spreadsheets as intensively as market participants, but in this case spreadsheets are used *for the purposes of monitoring, regulating and forecasting the markets themselves*. There is little evidence that regulators have put in place control processes to manage the spreadsheets they use to manage the financial system.

**3.5 Contagion**

Back in 2005, the anonymous UK regulatory interviewee clearly understood the relationship between Value at Risk (VAR), Tail Risk, Regulatory Capital & Financial Contagion. The following email was omitted from our earlier paper:

*"…it might be useful to give a sense of how much capital is at risk. Institutions measure market risk by VAR {Value at Risk}, which is to a 99% confidence level, daily horizon. Capital is set at between 10 and 20 times that number. 'Tail risk' is the risk*

*of extreme events, where all capital held by a firm is lost (and where potentially it is forced into default). This could be 40 times VAR or more, i.e. multiples of the capital buffer, and when this gets to such high levels, this has a 'contagion effect' – the firm cannot pay creditors out of capital, the creditors also realise losses, perhaps they too are forced into insolvency, thus forcing other firms, and so on. The total amount at risk in the global system is massive (obviously). How much of this is managed on spreadsheets? We don't know for sure, but we have specific examples where this happens. We also have a number of concrete examples where the spreadsheets (one due to a complex cell formula) failed. In one case a loss was realised. In another, there was a significant period when the risk was being mismanaged – fortunately this had the effect of reducing the risk, rather than increasing it. This risk impacts one of the FSA's 4 core objectives, which is market stability". [FSA, 2005a]*

The email was omitted because there was then no clear or definite link between spreadsheets and financial contagion. Plus of course, the risks of contagion were perceived to be small at the time. The fact that the UK regulator raised the issue of the relationship between financial contagion and spreadsheets is now an interesting point.

The author is aware of large spreadsheets incorporating monte carlo simulation techniques having been in widespread use in the past to calculate the Value at Risk of large firms. The UK regulator is also more recently aware of such use, expressing certain technical concerns:

*"We have come across some firms that perform Monte Carlo simulations for capital calculations using standard personal computer spreadsheet…this may prove inadequate when making a robust assessment of capital…". [FSA, 2008a]*

The degree to which it has been possible to calculate daily VAR on assets & liabilities where the primary transaction record is a spreadsheet is not known.

**4 THE RÔLE OF SPREADSHEETS IN THE COLLAPSE OF THE JAMAICAN BANKING SYSTEM**

The Jamaican Banking System collapsed in its entirety in the late 1990's partly due to the use of spreadsheets and a consequent failure to manage and control them [Lemieux, 2002, 2005].

The Jamaican banks relied heavily on spreadsheets because their major transaction processing and risk management systems failed to meet their management information and reporting requirements. Former employees of the failed banks revealed in later interviews that they had used spreadsheets in the following ways:

- Cash management
- Financial budgeting and control
- Analysis of product and customer profitability
- Analysis of the cost of capital
- Foreign Exchange management
- Credit decision management
- Interest rate sensitivity analysis
- Financial risk management
- Recording of proprietary transactions

The pattern of usage of spreadsheets in Jamaica in the late 1990's as outlined above is a small subset of the present pattern of usage of spreadsheets in the City of London and other major financial centres. The particular problems that gave rise to the Jamaican financial collapse were clear:

- Individualistic naming of files. These idiosyncratic file names obscured the content and purpose of the files, divorced the files from the business process and the people creating them (particularly if an individual had left the bank), prevented easy access and made management reporting very difficult.

- Ad hoc assignment of storage location. Important files were stored on personal hard drives without any document management, backup or archiving.

- No Document Retention policy. Data was deleted at the whim of the individual so that banks financial positions could not be calculated accurately or even at all.

- Failure to preserve a link to the business context. This rendered spreadsheets meaningless as a background to a particular business decision.

- Inability to guarantee the authenticity and reliability of spreadsheets. There were no controls over how spreadsheets were archived and no effort was made to "lock" down their content as part of a formal archiving process. As a result, their integrity was seriously questionable, since anyone could have changed the content in the intervening period and audit trail controls were weak to non-existent.

The problems with spreadsheets caused numerous problems, including:

*"Poor control over spreadsheets at Jamaican indigenous banks contributed to management information and external reporting problems (i.e., P&L distortions) that contributed to the banks' management and external regulators losing sight of the banks' true positions and exposures. This problem fed a downward spiral into liquidity crisis"[Lemieux, 2005]*

and

*"As the Jamaican financial crisis unfolded, the government also recognised that fraud and corruption had contributed to the collapse of indigenous banks. To address these allegations, it established a team of foreign and local forensic auditors to work with the police fraud squad to identify and take action on instances of fraud. Inaccessibility of source documents, however, seriously hampered the auditors' work. The work of reconstructing what in some cases were very convoluted financial transactions was made extremely difficult by the fact that critical records, many in spreadsheet form, were missing". [Lemieux, 2005]*

It is highly likely that these limitations in spreadsheet management and control experienced in Jamaica during the late 1990's are now being experienced in the wider, global economy.

## 5. UK REGULATOR'S KNOWLEDGE OF SPREADSHEET RISK

On four occasions during the period 2003-2007, an employee of the UK Financial Services Authority gave the keynote address at the annual conference of the European Spreadsheet Risks Interest Group [FSA, 2003, 2004, 2005, 2007].

The first presentation "*[End] User Computing in Financial Regulation*", outlined the FSA's approach. The FSA did not view End User Computing (EUC) as a bad thing, as it was essential to business efficiency:

> "The FSA is not, and never has been opposed to the use of spreadsheet and other user-developed applications for business critical purposes". [FSA, 2003]

It was clear however that the regulator expected appropriate controls to be in place. At that time it was clear that the senior management (of regulated firms) thought that end user computing was "bad" and sought to replace it with a *"strategic"* IT solution. Development of strategic solutions was hampered by the *"budget paradox"*. Generally, it was impossible to find a budget for any form of IT development required by the business, which implied that the firm *could not* afford it. However, it was always the case that some salaried employee of the firm could find the time, and a non-IT budget, to develop the required solution, which implied that the firm *could* afford it. This inevitably led to poorly managed solutions. It was clear to the FSA that EUC solutions were cheap to develop, but that the resulting disasters were not cheap. A change of mindset was called for, with attention being paid to training, standards and the traditional IT concerns of auditability, testability and maintainability. At the time, the view was that IT auditors focussed their attention on large information systems and tended to regard spreadsheets as being user problems, not of their concern.

The second presentation "*Why Banks use Spreadsheets*" sought to distinguish business critical applications from non business critical applications, identified the typical uses of spreadsheets illustrated by two case studies and concluded with a look of the cost of controls. The regulators position on the use of spreadsheets within business critical systems was clear:

> "It is crucial to distinguish so-called 'business critical' systems from all others. Financial regulators are primarily worried, for obvious reasons, about systems which are crucial to the successful and safe running of businesses. For example, systems which support the provisioning of liquidity to the money markets, which support global foreign exchange platforms, [or] which control the management of financial risk. This talk is concerned only with 'business critical' systems. I will talk about how many banks use business critical systems implemented on spreadsheets, based on evidence from many years of supervising banking systems". [FSA, 2004]

The regulator defined a business critical system as any system whose function is critical to the running of a business. Business critical systems were systems which: have more than one user; where the user probably doesn't know the "right answer"; where they were performing a cyclical or repetitive task; where there was usually data processing of some kind going on such as aggregation, enrichment, checking or MIS. Note that this definition of criticality differs from our own definition. Furthermore:

> *"Business critical applications are typically developed out of 'one off' spreadsheets which were not thrown away because they were seen to perform a useful routine function. Because the function was so routine and boring, they 'give' it to some less highly paid individual who does not understand what the spreadsheet is doing, with some instructions that s/he is able to follow only on a highly literal basis. [The] operator has little incentive to get the answer right apart from the fact they are paid to do it"* [FSA, 2004]

and

> *"Financial regulators see a lot of business critical applications developed on spreadsheets. They worry a lot about them, in case (a) they go wrong and the firm loses a lot of money in a way that disrupts the operation of the financial system. This has happened before, though not because of spreadsheet systems (so far) [and] (b) someone exploits the inherent security vulnerability of spreadsheets to commit some bad crime. This HAS happened before."* [FSA, 2004]

The reasons why end users build systems, rather than wait for IT to build them were clearly articulated:

> *"User developed systems, by contrast are built by small teams of people who understand exactly their own requirements, and are motivated to build exactly what they require in order to achieve their primary goal, which is a tool to further their business. Every minute they spend on the system (for them) is wasted effort, because they want to do something else. They only build the system because, without it, they cannot do what they actually want. Thus they have a strong incentive to minimise costs, and have by and large no reason to incur those costs in the first place."* [FSA, 2004]

The third and fourth presentations, in 2005 and 2007 both entitled *"Regulatory Update"* reviewed the progress on the Spreadsheet issue within regulated institutions over the preceding several years.

In the regulators view, the management mind set had changed as spreadsheets were increasingly accepted as strategic solutions, which was a major change from five years previously. Many firms had given up on the 'big solution' route and accepted end user computing for what it was. The regulator noted the increasing use of spreadsheet control solutions, which removed the burden of compliance from the end user.

The regulator noted some good news from the audit profession where they were seeing more mention of spreadsheets in audit reports and that increasingly EUC was part of the audit plan. The Sarbanes Oxley act had assisted in controlling EUC as it required firms to demonstrate that controls around financial reporting are adequate. This requires the auditor to assess the effectiveness of managements' assessment of internal controls. It was noted that the requirement to demonstrate that effective EUC controls are in place was particularly onerous because EUC applications are usually not documented. Most problems however, were still the result of poor use of EUC solutions. Good training was an obvious solution, but there was little budget.

## 6. CONCLUSIONS

We have shown in previous sections and the literature supporting them that:

   a) there is a large research base extending back over twenty years demonstrating that spreadsheets pose significant risks to organisations
   b) spreadsheets are integral to the function and operation of the global financial system and most users of the financial system do not realise this
   c) catastrophic collapse of one form or another due to the use of untested spreadsheets in the financial markets was predicted several years ago
   d) a recent previous regional banking collapse was partly caused by the use and abuse of spreadsheets
   e) the UK financial regulator was fully cognisant of the risks of using spreadsheets within the banking and other regulated financial sectors, the related possibility of contagion and relevant technical problems in the use of spreadsheet based monte carlo techniques to compute VAR based capital requirements
   f) credit derivatives were identified as being particularly reliant on spreadsheets
   g) a systemic collapse of the global financial system occurred during the period 2007-2009 where credit derivatives played a significant part in the destruction of capital

We have confidence in concluding that spreadsheets played a rôle, perhaps even a significant rôle, in the recent collapse of the financial system, affirming our research hypothesis. In our opinion, their primary rôle is centred around the fact that they were one of the principal technologies used in the Credit Derivatives marketplace. This market grew very quickly due to the ease with which it was possible to design and promulgate opaque financial instruments based on large spreadsheets with critical flaws in their key assumptions. Their secondary rôle is centred around their ubiquitous use such that many of the problems experienced in Jamaica – specifically the difficulty in valuing or risk assessing complex financial instruments pre and post crash contributed to the cause of the crash and the difficulties experienced in recovering from it. The tertiary rôle of spreadsheets is related to the issues associated with human error and other human factors, which are no doubt compounding difficulties in market recovery and will remain a problem for the future unless and until resolved.

During the course of the preparation of this article, we have identified a further set of risks associated with spreadsheets (and, by implication, other forms of end user computing). These give rise to the suspicion that the global financial system may be deeply flawed, requiring extensive modification and repair. Despite the recent recapitalisation and part nationalisation of the institutions, most of the spreadsheets still remain. In many cases these spreadsheets will now be operated by people not familiar with them and almost certainly outside of the domain where their assumptions and formulae were hoped to hold true. We are confident in predicting that further obvious problems with the financial system will come to light for some time unless and until untested and controlled spreadsheets play a less significant rôle in the financial system.

The ubiquitous use of uncontrolled spreadsheet technology leaves the financial system open to future systemic problems. Certain parts of the financial system have been homogenised by the ubiquitous use of spreadsheets. Those parts of the financial system are thus exposed to *all* the risks of spreadsheets, not merely the lack of testing or control. Note that the more heterogeneous interoperating systems developed using traditional IT methods which implement other parts of the financial system (eg cash transmission & retail banking), did not fail.

Arguably, if there had been a requirement that CDO's and such like had to be implemented or valued within traditional IT systems, or in spreadsheets using the full software development lifecycle, the CDO may never have seen the light of day.

**7 RECOMMENDATIONS**

Clearly, it is important that the financial system is resilient and stable enough to support the global economy. In our view, given the critical and burgeoning risks of spreadsheets and their ubiquity, this cannot happen without extensive changes in the demographics of their use & management [Powell, Baker, Lawson, 2005]. When we make a journey on public transport by air, rail, road or sea, there is high probability of our safe arrival. The same should apply to our expectations of surviving a financial journey, and much change will be required to achieve this.

Our previous recommendations still apply:

> *"As it is impossible to imagine the City of London operating without the widespread use of large spreadsheets, it is important to encourage firms to identify, prioritise, test, correct and control Key and Critical spreadsheets. Doing so will enable firms to improve profitability by avoiding the inevitable occurrence of large spreadsheet related mistakes, of which only the smallest and least embarrassing are being reported. Other actions that can be taken include: educating the market regarding the financial, regulatory and other risks of uncontrolled spreadsheet use; improving the regulatory framework regarding their use; promoting the emerging discipline of spreadsheet engineering; training management and staff to build & test more robust spreadsheet models; deploying software tools for detecting and correcting errors; deploying software tools to assist in the management of spreadsheets; encouraging the development of next generation spreadsheet technology and investigating other tools that have the advantages of spreadsheets but fewer or different disadvantages".[Croll, 2005]*

A very few organisations have voluntarily adopted Enterprise Spreadsheet Management systems, implementing them mainly on a departmental rather than a strategic basis. There is no evidence in the public domain or accessible to us that suggests that these control systems are truly effective during long-term use in the financial sector, though it is our hope and expectation that they are. We need to determine by detailed examination of the systems already in place that spreadsheet control systems do give the requisite degree of long term control to spreadsheet based financial systems, or financial systems where spreadsheets play a significant rôle.

Having determined that it is possible to effectively control spreadsheets, the regulators must compulsorily require that all participants in the global financial system discover, risk assess, test and control the use of spreadsheets already in place as part of the system. Spreadsheets that necessarily must become part of the financial system must go through the full software development lifecycle [Grossman, 2002]. They must be designed, implemented and controlled by people professionally trained in their creation, testing and deployment. This can only happen through global governmental and regulatory cooperation and may, due to cost, require a reduction in the scale and scope of the regulated financial system.

If it is the case that we cannot build a safe and effective global financial system using spreadsheets as a key component, then we will have no alternative but to gradually replace them, which will be time consuming and expensive, if indeed it is possible at all. The decision to replace a spreadsheet or set of spreadsheets with an alternative is a straightforward commercial decision, influenced by regulatory considerations, which would look at the risks, costs and benefits of each of the alternatives.

Finally, in our attempt to build a better financial system, it is very important to note that a century of research on human error informs us that humans have predictable limitations in many cognitive domains, including spreadsheets [Panko, 2007]. These limitations must be taken into consideration as we make decisions about technologies including spreadsheets. Panko's conclusions provide excellent strategic guidance, which we would be unwise to ignore:

> *"It is fundamentally important to understand that the problem is not spreadsheets. Asking what is wrong with spreadsheets has some value, but it is fundamentally the wrong issue. The real issue is that thinking is bad. Of course, thinking has some benefits, and we are accurate 96% to 98% of the time when we do think……The second fundamental thing to understand from human error research is that making large spreadsheets error free is theoretically and practically impossible….. {and to attempt to do so} is a testament to the profound human tendency to overestimate their ability to control their environment even when they cannot. The third fundamental thing to understand is that replacing spreadsheets with packages does not eliminate errors and many not even reduce them. The problem, again, is that thinking is bad. Unless there are no human decisions or design actions in using a package, there will be thinking, and therefore there will be errors"* [Panko, 2007]

We nevertheless hope that this paper has set out many of the hitherto largely unknown risks of the use of spreadsheets in the global financial system such that better informed decisions regarding their future use can be made. Whatever technology choices are made for the future, human factors [Thorne & Ball, 2005] will remain the overarching consideration.

## 8 ACKNOWLEDGEMENTS

The author thanks a number of EuSpRIG colleagues for comments on an early draft of this paper. This work was supported by the author during the period March-May 2009.There are no conflicts of interest to declare.